\documentclass[fleqn,10pt]{wlscirep}
\usepackage[utf8]{inputenc}
\usepackage[T1]{fontenc}
\usepackage{comment}
\usepackage{nameref}
\usepackage{hyperref}
\usepackage{longtable}
\usepackage{booktabs}
\usepackage{xcolor}
\usepackage{relsize}
\usepackage{soul}

\usepackage{lineno}

\title{Community detection in bipartite signed networks is highly dependent on parameter choice}	

\author[a,b]{Elena Candellone}
\author[a]{Erik-Jan van Kesteren}
\author[a]{Sofia Chelmi}
\author[a,b]{Javier Garcia-Bernardo}

\affil[a]{Department of Methodology \& Statistics, Utrecht University, Utrecht, The Netherlands}
\affil[b]{Centre for Complex Systems Studies, Utrecht University, Utrecht, The Netherlands}

\begin{abstract}
Decision-making processes often involve voting. Human interactions with exogenous entities such as legislations or products can be effectively modeled as
two-mode (bipartite) signed networks---where people can either vote positively, negatively, or abstain from voting on the entities.
Detecting communities in such networks could help us understand underlying properties: for example ideological camps or consumer preferences.
While community detection is an established practice separately for bipartite and signed networks, it remains largely unexplored in the case of bipartite signed networks. In this paper, we systematically evaluate the efficacy of community detection methods on projected bipartite signed networks using a synthetic benchmark and real-world datasets. Our findings reveal that when no communities are present in the data, these methods often recover spurious user communities. When communities are present, the algorithms exhibit promising performance, although their performance is highly susceptible to parameter choice. This indicates that researchers using community detection methods in the context of bipartite signed networks should not take the communities found at face value: it is essential to assess the robustness of parameter choices or perform domain-specific external validation.
\end{abstract}

\begin{document}
\maketitle

\section*{Introduction}\label{sec:intro}
Social groups often make decisions by voting. 
From politicians passing legislation, to online marketplaces where users review products, to social media users voting on popular content, these interactions reveal complex patterns, such as political affiliations, consumer preferences, or ideological camps. 
In such systems, individuals do not interact directly with each other, but with different entities (e.g., bills, products, articles) by voting for, against, or abstaining. 
These interactions can be modeled as bipartite signed networks, where two different types of nodes (bipartite) are connected through both positive and negative relationships (signed). 
A key insight that can be recovered from those networks is the presence of communities\cite{fortunatoCommunityDetectionGraphs2010}---groups of people with similar voting patterns.
These indicate alliances between politicians, hidden groups trying to promote products or articles, or ideological groups in the presence of polarization.\footnote{Instead of finding clusters of people with similar voting patterns, community detection methods could also be applied to find groups of entities (e.g., legislation, products or online content) that are voted similarly. In this paper, we focus only on clustering people.}
While many methods exist to analyze community structures in bipartite and signed networks separately \cite{barberModularityCommunityDetection2007, yenCommunityDetectionBipartite2020, beckettImprovedCommunityDetection2016, doreianPartitioningApproachStructural1996a, tomassoAdvancesScalingCommunity2022}, it is not clear whether they can handle the complexity of bipartite signed networks.

Community detection, a key research topic in network science\cite{fortunatoCommunityDetectionGraphs2010}, often aims to uncover structural patterns by identifying groups with dense intra-community connections and sparse inter-community connections. In the case of signed networks, it is not enough to have dense intra-community connections, but these connections should be positive, while negative connections should occur mainly between communities. Existing methods\cite{cucuringuSPONGEGeneralizedEigenproblem2019a, traagCommunityDetectionNetworks2009a, doreianPartitioningSignedSocial2009a} primarily focus on these principles of intra-community agreement and inter-community disagreement. These methods incorporate the information of negative relations in areas where it is crucial, such as finding communities in a network of friendships and enmities, to networks of alliances and wars\cite{nealSignTimesWeak2020,leskovecSignedNetworksSocial2010a,krosAvoidanceAntipathyAggression2021}. However, the structure of agreement and disagreement in bipartite networks presents a unique challenge: interactions occur between different sets of entities, while most methods assume a unique type.

\begin{figure}[h!]
    \centering
    \includegraphics[width=\textwidth]{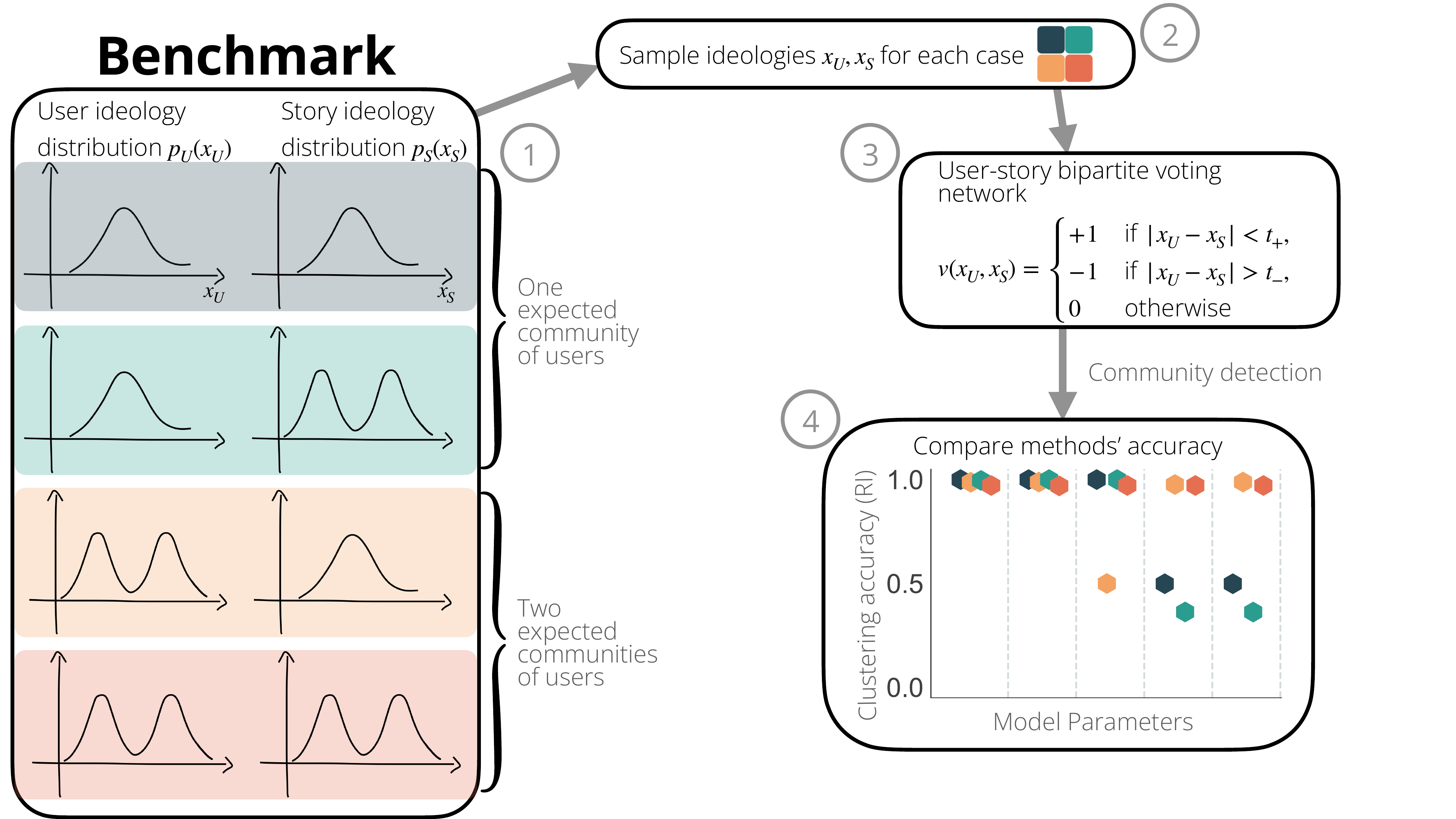}
    \caption{ \textbf{Synthetic bipartite signed networks.} We combine synthetic scenarios and insights from data to generate synthetic networks. Given a synthetic scenario, i.e., ideologies of users and stories modeled from probability distributions $p_{U}$ and $p_{S}$, which can be either unimodal or bimodal (step 1), we sample users' and stories' ideologies, $x_U$ and $x_S$, from those distributions (step 2). For each pair of user and story, the user then either votes positive, negative, or abstained from voting depending on the difference between the user and story ideologies (step 3), where the vote depends on two voting thresholds, $t_+$ and $t_-$, set to match the voting probability in real datasets. Given the bipartite network, we then project it into a unipartite network and apply the community detection methods for a wide range of parameter choices (step 4). }
    \label{fig:synth-mechanism}
\end{figure}

Here, we investigate the suitability of community detection methods devised for one-mode signed networks to capture different groups on bipartite signed networks. To address this question, we developed a benchmark (pictured in Figure \ref{fig:synth-mechanism}) to evaluate community detection methods. Our approach involves generating synthetic bipartite signed networks, where users (who may be in one or two communities in different scenarios) vote on articles, and project them into one-mode user-user networks. We then apply two well-established algorithms and systematically evaluate their ability to recover user-level communities. We use our methodology on two datasets: voting records from the US House of Representatives (1990-2022)\cite{OfficeClerkHouse} and user interactions on the Menéame platform\cite{MeneamePortadaNoticias}, a Reddit-like news aggregator.

Our findings reveal that community detection methods often identify spurious user communities not anticipated by the generative process, i.e., not following the initial user ideologies, and the results are susceptible to parameter choices. 

This paper is organized as follows. Section \hyperref[sec:background]{Background} contextualizes our contribution within the existing literature. The \hyperref[sec:methods]{Methods} section is subdivided into a description of the mathematical framework of bipartite signed networks (subsection \hyperref[subsec:bipartite-signed]{bipartite signed Networks: Mathematical Framework}), a detailed explanation of our benchmark (subsection \hyperref[subsec:synth-nets]{Generating synthetic bipartite signed networks}), and a contextualization of the community detection methods tested in this paper (subsection \hyperref[subsec:comm-det-methods]{Community Detection Methods}) along with the clustering evaluation metrics (subsection \hyperref[subsec:eval-measures]{Evaluating Community Similarity}). The \hyperref[sec:results]{Results} section demonstrates the effectiveness of the community detection methods on the synthetic and real-world datasets employed.

\section*{Background}\label{sec:background}
Community detection is an extensively researched problem in the field of network science \cite{fortunatoCommunityDetectionGraphs2010}. When dealing with unsigned networks, most techniques for detecting communities aim to maximize edge density within a community while minimizing it between communities. This interdisciplinary goal finds applications in various fields, such as biology (e.g., protein-protein interaction networks in cancer metastasis \cite{jonssonClusterAnalysisNetworks2006a}), social media studies (e.g., political blogs' ideological affiliations \cite{adamicPoliticalBlogosphere20042005}), and the science of science (e.g., scientists' collaboration network \cite{girvanCommunityStructureSocial2002}). However, signed networks provide additional information---the sign of the edge---that can be relevant for determining and interpreting communities in the network \cite{talagaPolarizationMultiscaleStructural2023b, bonchiDiscoveringPolarizedCommunities2019, tomassoAdvancesScalingCommunity2022}.

Research on signed networks is closely tied to the influential work of Heider \cite{heiderPsychologyInterpersonalRelations1958a} on Structural Balance Theory (SBT), later formalized by Harary and Cartwright \cite{cartwrightStructuralBalanceGeneralization1956}. SBT is based on the principles of ``the enemy of my enemy is my friend'' and ``the friend of my friend is my friend''. In network-theoretical terms, SBT posits that triangles with zero or two negative edges are balanced, while triangles with one negative edge are unbalanced. This concept extends to cycles (closed paths) of length $\geq 3$, stating that a cycle is balanced if the product of the edge signs is positive (i.e., there is an even number of negative interactions in a cycle). A network is considered strongly balanced if all cycles are balanced \cite{cartwrightStructuralBalanceGeneralization1956}.

SBT is regarded as one of the most relevant underlying mechanisms for the formation and evolution of signed networks, with various extensions that account for longer closed walks \cite{estradaWalkbasedMeasureBalance2014, arefMeasuringPartialBalance2018a, estradaRethinkingStructuralBalance2019a, talagaPolarizationMultiscaleStructural2023b} used to understand situations like alliances among countries \cite{doreianStructuralBalanceSigned2015,diaz-diazLocalBalanceSigned2023} and polarization on social media \cite{fraxanetUnpackingPolarizationAntagonism2023}. Methods for community detection in signed networks typically do not maximize structural balance directly but instead minimize a related metric, \textit{frustration}. Aref and Wilson \cite{arefBalanceFrustrationSigned2019} defined the \textit{frustration count} as the sum of the number of negative edges within a community and the number of positive edges between communities\cite{arefBalanceFrustrationSigned2019}. As expressed by Doreian and Mrvar \cite{doreianPartitioningSignedSocial2009a}, there is a parallelism between structural balance and frustration minimization. Two theorems by Cartwright and Harary \cite{cartwrightStructuralBalanceGeneralization1956} and by Davis \cite{davisClusteringStructuralBalance1967a} state that it is possible to find optimal partitions in a balanced network and that strong balance corresponds to null frustration (i.e., no frustrated edges).

While SBT and frustration provide a way to define communities in signed networks, applying them to bipartite signed networks has a key challenge: SBT and frustration are deeply linked to the presence of triangles, while bipartite networks have no cycles of odd length \cite{asratianBipartiteGraphsTheir1998}. Bipartite networks exhibit disassortative properties, with two sets of nodes forming edges only across different sets. Often, the two sets of nodes have different types, such as movies and actors \cite{wattsCollectiveDynamicsSmallworld1998} or buyers and products on an online marketplace \cite{derrBalanceSignedBipartite2019}. A potential solution to the missing interpretability of SBT on bipartite networks is to create a one-mode network where two nodes interact if they are connected to the same entity in the bipartite network. For example, this involves connecting politicians who voted on the same bill.  

A body of literature within political science has traditionally dealt with a specific type of signed bipartite networks: co-voting networks of the United States Senate and House of Representatives \cite{leeVaryingcoefficientModelsDynamic2020a, andrisRisePartisanshipSuperCooperators2015}. The goal of that body of literature is to estimate legislators' ideology, for which they use ideal point models, where the probability of voting positive versus voting negative or abstaining is modeled as a function of a latent difference in ideology between the legislator and the bill  \cite{enelowSpatialTheoryVoting1984, pooleSpatialModelLegislative1985,pooleSpatialModelsParliamentary2005, clintonStatisticalAnalysisRoll2004, yuSpatialVotingModels2021}. These models combine methods from economics (spatial models) with measurement theory from psychometrics~\cite{carrollIdealPointEstimation2023}, based on the assumption that latent ideologies can explain human behaviors such as voting.
The US co-voting networks have also been used to measure ideology \cite{moodyPortraitPoliticalParty2013, waughPartyPolarizationCongress2009}, calculate community structure \cite{zhangCommunityStructureCongressional2008,muchaCommunityStructureTimeDependent2010}, points in which voting dynamics change \cite{wilsonModelingDetectingChange2016} or to predict the sign of the votes \cite{derrBalanceSignedBipartite2019,huangSignedBipartiteGraph2021}. However, given the dense nature of voting interactions (i.e., most legislators would vote to all the roll calls), these methods consider only a binary choice between agreement and disagreement---i.e., disagreement encapsulates both negative and absence of interactions. 

Our work complements this body of literature by extending the study of voting patterns to sparse (offline) voting records, allowing us to understand how network density affects community detection. We develop a generative model closely related to the ideal points model literature~\cite{pooleSpatialModelLegislative1985, pooleSpatialModelsParliamentary2005, clintonStatisticalAnalysisRoll2004, barberaBirdsSameFeather2015}. We aim to incorporate the interplay between voting patterns and the intrinsic ideologies of the voters, based on the assumption that human voting tendencies are influenced by their internal belief systems.

\section*{Methods}\label{sec:methods}

\subsection*{Bipartite Signed Networks: Mathematical Framework}\label{subsec:bipartite-signed}
Consider a network $G = (V, E)$, where $V$ is the set of nodes of size $N$, and $E$ is the set of edges of size $M$. In a bipartite network, nodes are divided into two disjoint sets, which we label $U$ and $S$ to represent users and stories (or legislators and bills)\footnote{$V = U \cup S$ and $U \cap S = \emptyset$.}. The edges connect nodes from one subset to the other only, i.e., $E \subseteq U \times S$. A network can be uniquely represented by an adjacency matrix $A$ of dimension $N \times N$, with each element given by the edge weight such that $A_{i,j} = \omega_{i,j}$ if the edge $(i,j) \in E$, else $A_{i,j} = 0$. The edge weight, in the case of signed networks, can be either binary (negative or positive, $\omega: E \rightarrow [-1,1]$) or real ($\omega: E \rightarrow \mathbb{R}$). In this paper, we consider the edge weights as simplified to the binary choice. As a special property of undirected bipartite networks, the adjacency matrix can be written as
\begin{equation*}
A = \begin{pmatrix}
0 & B\\
B^T & 0
\end{pmatrix},
\end{equation*}
by ordering the nodes such that $\{v_1, \dots, v_{|U|}\} \in U$ and $\{v_{|U|+1}, \dots, v_N\} \in S$. The matrix $B$ is called the incidence matrix (or biadjacency matrix) and has dimension $|U|\times |S|$. The elements of $B$ represent users' (in rows) votes on stories (columns).

From the incidence matrix, it is possible to obtain the bipartite projection of the network, denoted as $P = B \cdot B^T$. In the bipartite projection, nodes correspond to those from set U, and edges represent the presence of a common connection to nodes in set S.
In the case of a weighted bipartite network, the edge weight in the bipartite projection is the sum of the edge weights
\begin{equation}\label{eq:weight-projection}
 \omega'_{kl} = \mathlarger{\mathlarger{\sum}}_{z \text{ s.t. } (k,z) \text{ and } (l,z) \in E} \omega_{k,z} \; \omega_{z,l}.
\end{equation} 

To provide intuition, if two politicians always agree on their votes, the edge between them has a positive weight, representing agreement. Conversely, if they always vote differently, their edge has a negative weight, representing disagreement. We project the bipartite signed network to analyze the performance of the community detection methods designed for unipartite networks, where all nodes are of the same type (e.g., we aim to find communities of users on social media or members of the US Congress). One reason for this choice is the absence of SBT-related community detection methods designed specifically for bipartite signed networks. While, for the methods that can accommodate bipartite structures, we use the adjacency matrix. In the following, we provide a detailed description of the generative mechanism to create synthetic bipartite signed networks.

\subsection*{Generating synthetic bipartite signed networks}\label{subsec:synth-nets}
A body of literature within political science has traditionally focused on a specific type of signed bipartite network: co-voting networks of the United States Senate and House of Representatives~\cite{leeVaryingcoefficientModelsDynamic2020a, andrisRisePartisanshipSuperCooperators2015}. The main goal of this literature is to estimate the ideology of legislators using ideal point models, which model the probability of individual voting in favor of a bill based primarily on the difference between the latent ideologies of the legislator and the bill~\cite{enelowSpatialTheoryVoting1984, pooleSpatialModelLegislative1985, pooleSpatialModelsParliamentary2005, clintonStatisticalAnalysisRoll2004, yuSpatialVotingModels2021}. While these models are widely used to infer ideological positions in contexts such as the US Congress~\cite{pooleSpatialModelsParliamentary2005, clintonStatisticalAnalysisRoll2004}, they could also be used as generative models, i.e., to generate votes, given the parameters of the model and ideological positions of individuals and bills. However, given the dense nature of voting interactions, these models equate abstentions and votes against, and thus cannot generate sparse voting data---a common characteristic in social media. While spatial models can be extended to treat abstentions separately~\cite{kubinecGeneralizedIdealPoint2019}, the interpretation of the results depends on parameters that are difficult to set and interpret. To address these limitations, we use an interpretable threshold-based generative model that explicitly incorporates three outcomes: abstaining, voting yes, or voting no.

To assess the performance of current community detection methods, we generate synthetic networks with and without polarized users, with realistic voting mechanisms. We consider four scenarios that entail different choices for the \textit{user} and \textit{story ideologies}. In two scenarios, the ideology of the users is drawn from a Gaussian distribution. Therefore, we expect the community detection algorithms to find one community. In the other two scenarios, the ideology of the users is polarized (drawn from a mixture of two Gaussian distributions), and we expect the community detection algorithms to find two communities. The generative process is illustrated in Figure \ref{fig:synth-mechanism} and the following paragraphs.

\begin{enumerate}
    \item \textbf{Sampling ideologies.} We assign a latent \textit{ideology} score to each of $|U|$ \textit{users} and $|S|$ \textit{stories}, representing the two types of nodes in the bipartite signed network. While we focus our example on users voting on news stories for simplicity, this latent score could represent the ideology of politicians and bills, or the type of consumer preferences and product characteristics (e.g. gelato vs ice-cream). For each type of entity ($U$ and $S$) we sample the ideology score from either one Gaussian distribution or a mixture of two Gaussian distributions with given averages ($\mu_i$) and standard deviations ($\sigma_i$). Specifically,
        \begin{align}\label{eq:ideology_distributions}
        \begin{split}
        x_U \sim p_U(x_U) &= \pi_U \cdot \mathcal{N}(\mu_1,\,\sigma_1^{2}) + \left( 1- \pi_U \right) \cdot \mathcal{N}(\mu_2,\,\sigma_2^{2}),\\
        x_S \sim p_S(x_S) &= \pi_S \cdot \mathcal{N}(\mu_3,\,\sigma_3^{2}) + \left( 1 - \pi_S \right) \cdot \mathcal{N}(\mu_4,\,\sigma_4^{2}),
        \end{split}
        \end{align}
        where $p_U(x_U)$ and $p_S(x_S)$ are respectively the probability distributions of \textit{users' and stories' ideologies}. When the ideology is sampled from \textit{one} Gaussian distribution, $\pi_U = \pi_S = 1$. When is sampled from a mixture of two, we set $\pi_U = \pi_S = 0.5$, which means that we assume that there is no imbalance in the sizes of two groups of \textit{ideologies}. We set $\sigma_1 = \sigma_2 = \sigma_3 = \sigma_4 = 0.1$ and  $\mu$ depending on scenario's choice.
        The four scenarios,  illustrated in Figure \ref{fig:synth-mechanism}A, are:

\begin{itemize}
    \item \textit{Users and stories not polarized} (\textbf{U NP S NP}): we consider both \textit{users} and \textit{stories} as not polarized. In other words, both \textit{ideologies} are drawn from one Gaussian-distribution ($\mu_1 = \mu_2 = \mu_3 = \mu_4 = 0$ in eq. \ref{eq:ideology_distributions}). We expect to find all \textit{users} in the same community.

    \item \textit{Users not polarized, stories polarized} (\textbf{U NP S P}): Users are not polarized, and stories are polarized. In other words, \textit{users}' ideologies are drawn from one Gaussian distribution ($\mu_1 = \mu_2 = 0$) and \textit{stories}' idelogies are drawn from a mixture of two Gaussian distributions ($\mu_3 = -\mu_4 = 1$). We expect to find all \textit{users} in one community.
    
    \item \textit{Users polarized, stories not polarized} (\textbf{U P S NP}): Users are polarized, and stories are not polarized. In other words, $\mu_1 = - \mu_2 = 1$ and $\mu_3 = \mu_4 = 0$. We expect to find all \textit{users} in two communities.
    
    \item \textit{Users and stories polarized} (\textbf{U P S P}): We consider both \textit{users} and \textit{stories} as polarized. In other words, $\mu_1 = -\mu_2 = 1$ and $\mu_3 = -\mu_4 = 1$. We expect to find all \textit{users} in two communities.
    
\end{itemize}
        
        After this step, every node has an \textit{ideology} assigned. Users will vote positive to a story if the difference of their ideologies is below a threshold, negative if it is above a threshold, and abstain otherwise. 
    \item \textbf{Computing the ideology difference distribution.} If two random variables are Gaussian-distributed, then their linear combination is also Gaussian-distributed. For a detailed proof, please refer to the Appendix \hyperref[appendix:ideology-diff]{Ideology Difference}. The \textit{ideology difference} between users and stories is distributed as a Gaussian Mixture with at most four peaks, e.g., in the case when all averages are different: 
    \begin{align}\label{eq:ideology-difference}
        \begin{split}
        x_U-x_S \sim p_{\text{diff}}(x_U-x_S) &= \pi_U \cdot \pi_S \cdot \mathcal{N}(\mu_1-\mu_3,\,\sigma_1^{2}+\sigma_3^{2})+\pi_U \cdot \left( 1 - \pi_S \right) \cdot \mathcal{N}(\mu_1-\mu_4,\,\sigma_1^{2}+\sigma_4^{2})+\\
        &+\left( 1- \pi_U \right) \cdot \pi_S \cdot \mathcal{N}(\mu_2-\mu_3,\,\sigma_2^{2}+\sigma_3^{2})+ \left( 1- \pi_U \right) \cdot \left( 1 - \pi_S \right) \cdot \mathcal{N}(\mu_2-\mu_4,\,\sigma_2^{2}+\sigma_4^{2}).
        \end{split}
        \end{align}
    \item \textbf{Extracting voting probabilities from data.} The thresholds determining the type of vote (positive, abstain, or negative) are calculated so the voting probability in the simulation matches the voting probability in the real data (Section \hyperref[subsec:datasets]{Datasets}). In the simplest case, we define constant voting probabilities extracted from data. Voting probabilities $v_+$ ($v_-$) are computed as the average number of positive (negative) votes per user, normalized by the number of stories. We also consider a more sophisticated alternative (Appendix \hyperref[appendix:deg-corr]{Degree-corrected synthetic networks}), where we assign different voting probabilities to each user, given the distribution of voting probabilities in the real data's bipartite network.
    \item \textbf{Defining voting thresholds.} We determine voting thresholds, denoted as $t_+$ and $t_-$, that meet the following conditions:
        \begin{align}
            \int_{-t_+}^{t_+} p_{\text{diff}}(x) \,dx = v_+ \;\;\;\; \text{   and   } \;\;\;\;
             \int_{t_-}^{+\infty} p_{\text{diff}}(x) \,dx = \frac{v_-}{2}.
        \end{align}
        where $p_{\text{diff}}$ is the ideology difference distribution given in Equation \ref{eq:ideology-difference}, $v_+$ and $v_-$ are the voting probabilities obtained from data. In this way, we have an amount of positive and negative votes proportional to what is expected from real data.
    \item \textbf{Creating the bipartite signed  network.} Finally, for each \textit{user} and \textit{story} with \textit{ideologies} $x_U$ and $x_S$, the vote of user $U$ to story $S$ is determined by
        \begin{equation}
        v(x_U,x_S) =
        \begin{cases}
        +1 &\text{ if } |x_U-x_S| < t_+,\\
        -1 &\text{ if } |x_U-x_S| > t_-,\\
        0 &\text{ otherwise}.
        \end{cases}
        \end{equation}
    After this step, we obtain a bipartite signed network, where the voting patterns respect both controlled, synthetic settings, and real data properties.
\end{enumerate}
Finally, given the incidence matrix of the synthetic network, we obtain the bipartite projection where the nodes are the \textit{users} and the edge weight is given by Equation~\ref{eq:weight-projection}.

The generative approach allows us to generate synthetic bipartite signed networks with voting probabilities matching real datasets, but where we modify the number of clusters expected (either one or two). 

\subsection*{Community Detection Methods}\label{subsec:comm-det-methods}
We compare two widely used community detection methods: SPONGE and community-spinglass. Both methods infer communities by minimizing network frustration---that is, by assigning nodes into communities in a way that minimizes the number of positive edges and maximizes the number of negative edges between communities. Since this problem is  NP-hard \cite{arefBalanceFrustrationSigned2019}, various approximations have been proposed. We chose SPONGE and community-spinglass because they represent two different approaches to approximating frustration minimization: SPONGE is a spectral method (i.e., using the spectrum of the Laplacian matrix to infer the community structure), while community-spinglass is a method that optimizes a quality function (i.e., finding the ground state of the Hamiltonian defined in Eq.~\ref{eq:hamiltonian}). Both SPONGE and community-spinglass are designed for one-mode (unipartite) networks. As an additional test, we also analyzed an algorithm that can in principle handle bipartite networks: the Stochastic Block Model (SBM). For more details and results on the SBM, please refer to Appendix \hyperref[subsubsec:sbm]{Weighted SBM with Edge Attributes}.

\subsubsection*{SPONGE}\label{subsubsec:sponge}

The \textit{SPONGE} method (Signed Positive Over Negative Generalized Eigenproblem), introduced by Cucuringu et al. \cite{cucuringuSPONGEGeneralizedEigenproblem2019a} minimizes the number of ``violations'', which is equivalent to frustration minimization. These violations consist of positive edges between communities and negative edges within communities. The method achieves this objective by incorporating a regularization term for the size of the partitions. This regularization term avoids setting each node in its community.

The optimization problem is formulated as follows:
\begin{equation}\label{eq:hamiltonian}
    \min_{C_1, \dots, C_k} \sum_{i=1}^k \frac{x_{C_i}^T (L^+ + \tau^- D^-) x_{C_i}}{x_{C_i}^T (L^- + \tau^+ D^+) x_{C_i}},
\end{equation}
where $\{C_1, \dots, C_k\}$ represents the communities, $L^+$ and $L^-$ are the Laplacian matrices of the network with only positive and negative edges, respectively, and $D^+$ and $D^-$ are diagonal matrices. $x_{C_i}$ with $C_i \in \{C_1, \dots, C_k\}$ are vectors where $(x_{C_i})_j =1$ if node $j \in C_i$, and 0 otherwise. The latter terms ($\tau \; D$) correspond to the previously mentioned regularization for the partition size, while the Laplacian acts as a \textit{frustration count}. This optimization problem is equivalent to a spectral problem, as discussed in detail in the original paper \cite{cucuringuSPONGEGeneralizedEigenproblem2019a}. An open-source Python implementation of the \textit{SPONGE} algorithm is available on GitHub \cite{alanturinginstituteSigNetPackage2023}.

The number of clusters $k$ is a required parameter, while default values for the regularization parameters are set to $\tau^+ = \tau^- = 1$. We conducted tests for different values of $k$, ranging from $k=1$ (no communities) to $k=10$. We chose several values of $k$ to quantify the impact of this choice on the identification of the expected communities. To account for the stochastic nature of the method, we repeated each iteration ten times for every value of $k$.

\subsubsection*{community-spinglass}\label{subsubsec:spinglass}

The \textit{community-spinglass} method, initially proposed by Reichardt and Bornholdt for unsigned networks \cite{reichardtStatisticalMechanicsCommunity2006}, presents a community detection approach grounded in statistical mechanics. Traag and Bruggeman later extended this method to signed networks \cite{traagCommunityDetectionNetworks2009a}.

The method is based on finding a Hamiltonian's ground state (i.e., minimal energy), defined as a function of the partition of nodes into communities $\{\sigma\}$. The final form of the Hamiltonian, detailed in the original paper by Traag and Bruggeman \cite{traagCommunityDetectionNetworks2009a}, is given by:
\begin{equation}
    H\left(\{\sigma\}\right) = - \sum_{ij} \left[A_{ij} - \left(\gamma^+ p_{ij}^+ - \gamma^- p_{ij}^-\right)\right] \delta\left(\sigma_i, \sigma_j\right),
\end{equation}
Here, $A$ is the adjacency matrix (could be both the adjacency matrix previously defined as $A$, or, as in our case, the bipartite projection matrix $P$), and $\gamma^+$ (or $\gamma^-$) is a parameter rewarding the presence (or absence) of positive (or negative) links within a community. The term $\delta\left(\sigma_i, \sigma_j\right)$ equals $1$ when nodes $i$ and $j$ are part of the same community ($\sigma_i == \sigma_j$), and thus counts only interactions found in the same community. If $\gamma = 1$, present positive and missing negative edges are given equal importance. The case $\gamma^+ = \gamma^- = 0$ corresponds to frustration minimization. The parameters $p_{ij}$ represent the expected value for the link between two nodes in a random null network.
There is an open-source implementation of this algorithm in the Python version of the \textit{igraph} package \cite{csardiIgraphSoftwarePackage2005}. To determine the sensitivity of the results to the $\gamma$ parameters (the one that recovers the real community structure in our synthetic scenarios), we conducted tests with combinations of $\{ \gamma^+, \gamma^- \} \in \{0.5, 1, 2\}$. Due to the time-consuming nature of the computations, we did not repeat each measure multiple times, despite the stochastic nature of the methods.

\subsection*{Evaluating Community Similarity}\label{subsec:eval-measures}

To assess how well the different community detection methods capture the community structure, we use the \textit{Rand Index} \cite{randObjectiveCriteriaEvaluation1971} implemented in the Python \textit{scikit-learn} package \cite{scikit-learn}. Note that we expect robust methods to identify only one community when there is no meaningful structure in the network. Since we draw the user ideology from a mixture of Gaussian distributions, we know which users belong to the same community. Given this true community labeling and the estimated labeling using the community detection algorithms,  the Rand Index is defined as:

\begin{equation}
    \text{RI} = \frac{\text{TP} + \text{TN}}{\binom{n}{2}},
\end{equation}

where $\text{TP}$ ($\text{TN}$) is the number of pairs of elements found in the same (different) community in the true and estimated community labeling. The denominator accounts for all possible combinations of pairs between the nodes. The Rand Index is bounded between 0 (indicating that every pair of users is clustered in the wrong community) and 1 (indicating a perfect matching). We focus mostly on the Rand Index since it produces interpretable estimates even when only one community is present in the data. We also consider other evaluation metrics, but they fail to account for situations where only one community is expected (see Appendix \hyperref[appendix:eval-metrics]{Clustering evaluation metrics}).

\subsection*{Datasets}\label{subsec:datasets}
In addition to evaluating the community detection methods on simulated network data, we apply and compare them on two real-world datasets: the Menéame votes and the US House of Representatives votes. These examples represent two different interaction modalities: sparse interactions, typical of the online world, and dense interactions, typical of the offline world.

\subsubsection*{Sparse voting: Menéame social platform}\label{subsubsec:meneame-data}
We gathered data from the Spanish social media platform known as Menéame \cite{MeneamePortadaNoticias}. In this paper, each post is referred to as a \textit{story} and includes a hyperlink to a news article and a summary provided by the user sharing the story. Users can engage with the story by liking or disliking them. Positive votes were encoded for each story as $+1$ and negative votes as $-1$. We constructed an incidence matrix $B_{M}$ where the rows represent users, and the columns represent the stories on the homepage. We collected data from 11,628 users and 48,561 stories, covering the period from the 27th of November 2022 to the 17th of July 2023. The bipartite projection network comprises 11,628 nodes and 9,834,491 edges. The voting probabilities are $v_+^M=1.30741\%$ and $v_-^M =  0.15602\%$ respectively for liking and disliking. We use these probabilities to create synthetic networks (Section~\hyperref[subsec:synth-nets]{Generating synthetic bipartite signed networks}) with 1,000 users and 4,176 stories. The number of stories is set to match the proportion of stories/users found in the data. We find that there are 2.4 \% of cases where couples of users have the same number of agreements and disagreements, resulting in the absence of an edge between them. 

\subsubsection*{Dense voting: House of Representatives}\label{subsubsec:clerk-data}
We obtained data from the Clerk of the House of Representatives website\cite{OfficeClerkHouse}. For each bill, we retrieved the following information: the name of the Representative, their party, and their vote. Each Representative could cast a vote of \textit{Aye} (previously \textit{Yea}), which was encoded as a positive vote with a weight of $+1$, \textit{Nay} (previously \textit{No}), encoded as a negative vote with a weight of $-1$, or be categorized as \textit{Present} or \textit{Not Voting}, both encoded as a null vote with a weight of $0$. From this, we created an incidence matrix $B_{H}$ of the bipartite network, where the rows are the Representatives and the columns are the Bills. We collected data from 1990 to 2022, encompassing votes from 1801 members of the House of Representatives on 20385 different Bills. Table \ref{tab:clerk-descr-stats} in the Appendix \hyperref[appendix:descr-stat]{Descriptive statistics of co-voting networks of House of Representatives} presents basic descriptive statistics for each snapshot network. The average number of nodes per snapshot is $N = 441$ with a standard deviation of $SD = 4$. The average number of edges is $M = 96726$ with a standard deviation of $SD = 1653$. We computed voting probabilities for each snapshot, and the average positive voting probability is $v^{H}_+ = 60.09 \%$ with standard deviation $SD = 4.82 \% $, while the average negative voting probability $v^{H}_- = 33.14 \%$ with standard deviation $SD = 5.14 \%$.
 
For both datasets, we then obtained the (dis)agreement unipartite network by multiplying the incidence matrix by its transpose, resulting in $P_{H} = B \; B^T$ (as defined in Subsection \hyperref[subsec:bipartite-signed]{Bipartite Signed  Networks: Mathematical Framework}). This projection is used for the \textit{SPONGE} and \textit{community-spinglass algorithms}.

\section*{Results and Discussion}\label{sec:results}
\subsection*{Sparse (social) networks}
\subsubsection*{Evaluation on synthetic networks}
\begin{figure}[h!]
    \centering
\includegraphics[width=\textwidth]{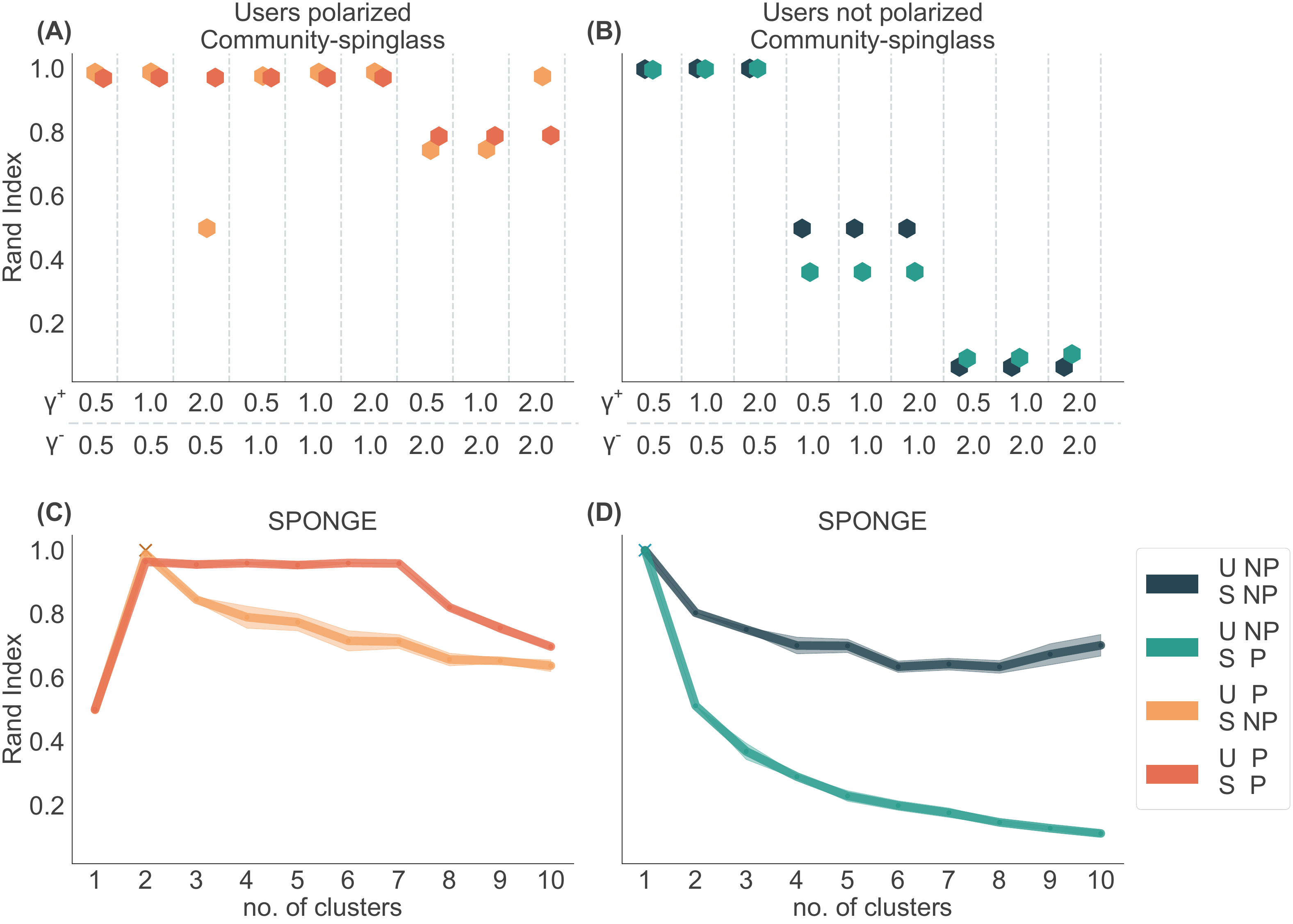}
    \caption{
    \textbf{Community Detection on Sparse Synthetic Networks.} 
    We tested the community detection methods on the four synthetic scenarios. Panels (A,C) show the results for the scenarios with two communities of users, while panels (B,D) show the results with one community.  We used the Rand Index to evaluate the performance of the algorithms with different parameter choices.  Higher values of the Rand Index indicate better alignment between expected and empirical communities. Panels (A-B) show the results for \textit{community-spinglass}. We experimented with different combinations of the parameters $\{\gamma^+, \gamma^-\} \in \{0.5, 1, 2\}$. Lower values of $\gamma^+$ indicate less importance given to positive ties in a community, whereas lower values of $\gamma^-$ penalize the presence of negative links within a community.
     Note that $\gamma^- = 0.5$ is generally the best parameter choice, as it finds the expected communities for the synthetic scenarios. Panels (C--D) show the results for \textit{SPONGE}. We conducted tests for different values of the number of clusters $k$, ranging from $k=1$ (no communities) to $k=10$, and repeated ten times each iteration, due to the stochasticity of the model. Error bars show the standard deviation around the mean. Note that the algorithm correctly identifies the expected communities in scenarios with polarized users (panel C), while it generates spurious communities in cases where stories introduce a latent ideology to a crowd of neutral users (the \textbf{U NP S P} case in panel D). 
}
    \label{fig:uniform-red}
\end{figure}

We first tested the performance of the algorithms on a sparse network, similar to those observed in online social networks. Figure \ref{fig:uniform-red} shows the results using our preferred measure for performance (the Rand Index), while Figure \ref{fig:uniform} in the Appendix shows that the results hold for other similarity measures.
When analyzing the scenarios in which the users' ideology was drawn from a mixture of two Gaussian distributions (and thus two ideological communities of users are expected), we found that \textit{community-spinglass} and \textit{SPONGE} had a Rand index over 75\%---i.e., more than 75\% of user pairs are clustered in the right community (Figure~\ref{fig:uniform-red}A,C). The constant performance of \textit{SPONGE} indicates that when enforcing a higher number of communities than expected, the algorithm identifies those communities by adding a few users to the remaining communities. \textit{community-spinglass} performed better when the negative votes within a community were penalized more (i.e. when $\gamma^-$ is low or medium)---i.e., with parameters matching the principle of frustration minimization.

The results were more diverse for the scenarios in which the users' ideology was drawn from one Gaussian distribution (and thus one community is expected, Figure~\ref{fig:uniform-red}B,D). 
Similarly to the previous case, the \textit{community-spinglass} method correctly recognizes all expected scenarios when negative votes within a community are penalized heavily ($\gamma^- = 0.5$), while the results were independent of the weight of positive links within a community ($\gamma^+$). 
\textit{SPONGE} performs relatively well when neither the users nor the stories are polarized---Rand index over 70\% even for a high number of clusters, but not when the stories are polarized. In the latter case, increasing the fixed number of clusters drastically reduces the Rand Index value, likely due to the latent ideology of the stories that split the group of users into those that are closer to one type of story and those that are closer to the other (even when the users are not very different from each other).

Finally, we allowed different users to have different voting probabilities, finding similar results (see Figure \ref{fig:deg-corr} in Appendix \hyperref[appendix:deg-corr]{Degree-corrected synthetic networks}) except for the scenario where neither users nor stories are polarized. In this scenario, SPONGE recovers the correct results (one community of users), while \textit{community-spinglass} is only able to recover the correct results when both negative votes within a community, and positive links within a community are weighted more ($\gamma^- = 0.5, \gamma^+ = 2$).

\subsubsection*{Evaluation on real networks: Menéame data}\label{subsubsec:meneame-data-results}
We applied the community detection methods on the Meneame co-voting network. We find that both \textit{community-spinglass} and \textit{SPONGE} find two or three large communities (depending on the parameter choice) and several small communities.

Table \ref{table:spinglass-meneame} presents the results for \textit{community-spinglass}. We restricted $\gamma^+$ and $\gamma^-$ to $\{0.5, 1\}$, the choices giving best results on synthetic networks. The algorithm identifies one larger community, followed by another of less than 1000 users when $\gamma^+ = 0.5$ is selected. Setting $\gamma^+ = 1$ results in three main communities. Therefore, by emphasizing the importance of negative ties while treating positive links as neutrally relevant, we observe a structural tripartition of this empirical network. 

\begin{longtable}[c]{@{}ll|lllllllllllllllllll@{}}
\toprule
\multicolumn{2}{l|}{community id.} & 0    & 1    & 2    & 3  & 4  & 5  & 6 & 7 & 8 & 9 & 10 & 11 & 12 & 13 & 14 & 15 & 16 & 17 & 18 \\
\endhead
\bottomrule
\endfoot
\endlastfoot
$\gamma^+$      & $\gamma^-$     &      &      &      &    &    &    &   &   &   &   &    &    &    \\* \midrule
0.5             & 0.5          &  10629 & 978 & 8 & 5 & 1 & 1 & 1 & 1 & 1 & 1 & 1 & 1 &  &  &  &  &  &  &  \\
0.5             & 1.0            &  10756 & 858 & 8 & 2 & 1 & 1 & 1 & 1 &  &  &  &  &  &  &  &  &  &  &  \\
1.0             & 0.5            &6166 & 2902 & 2323 & 121 & 39 & 28 & 17 & 10 & 6 & 3 & 3 & 2 & 2 & 1 & 1 & 1 & 1 & 1 & 1  \\
1.0             & 1.0            & 6059 & 3002 & 2530 & 13 & 9 & 6 & 3 & 2 & 2 & 1 & 1 &  &  &  &  &  &  &  &   \\* \bottomrule
\caption{
 \textbf{Community-Spinglass on Menéame Network.} We tested the \textit{community-spinglass} algorithm on the co-voting network extracted from the social media platform Menéame. We used the parameter choices found most suitable on synthetic networks (see Figure \ref{fig:uniform-red}A--B). When $\gamma^+ = 0.5$, we observe the presence of a single prominent community, followed by a secondary one with less than 1000 users. In contrast, with $\gamma^+ = 1$, the users are divided into three main communities, with approximately 6000, 3000, and 2000 users.}

\label{table:spinglass-meneame}\\
\end{longtable}
Next, we applied the \textit{SPONGE} algorithm on real data, as highlighted in Table \ref{table:sponge-meneame}. Our findings indicate that by increasing the fixed number of clusters, a larger community of approximately 7000 users can be identified. Additionally, we observed that a community initially consisting of over 4400 users tended to bifurcate. It resulted in two distinct communities, one with around 3400 users and the other with approximately 680 users.

Since the Menéame data does not contain information about the political leanings of the users, we compare the overlap between users for the parameter choices when both algorithms find either two or three communities (Appendix \hyperref[appendix:comparison-real-data-communities]{Comparing algorithms on real data}, Figure \ref{fig:methods-comparison-meneame-real}). In the case of two communities, there is no overlap between both methods. Both communities of \textit{SPONGE}, of  7,087 and 4,396 users, get mapped to the largest community of \textit{community-spinglass}, of 10,629 users. 

We find considerable overlap in the case of three communities. 88--89\% of the users in the two largest communities of \textit{community-spinglass} are mapped to the two largest communities of \textit{SPONGE}, while all users in the smallest community of \textit{community-spinglass} (683 users) are mapped to the smallest community of \textit{SPONGE} (2,323 users, the rest being mapped to the largest community of \textit{community-spinglass}). Our results highlight the sensitivity of the results to parameter choice, and stress the need for more robust assessment and external validation when dealing with real data.

\begin{longtable}[c]{@{}lllllllllll@{}}
\toprule
\begin{tabular}[c]{@{}l@{}}community id.\\ no. of clusters\end{tabular} & 0     & 1    & 2    & 3    & 4    & 5    & 6   & 7   & 8    & 9   \\* \midrule
\endhead
\bottomrule
\endfoot
\endlastfoot
1                                                       & 11628 &     &    &     &     &     &    &    &     &    \\
2                                                       & 11591 & 37   &     &     &     &     &    &    &     &    \\
3                                                       & 11583 & 37    & 8   &     &     &     &    &    &     &    \\
4                                                       & 7087  & 4496   & 37 & 8    &     &     &    &    &     &    \\
5                                                       & 7172  & 4410    & 25   & 13   & 8 &     &    &    &     &    \\
6                                                       & 7247  & 4325   & 25    & 13   & 10   & 8 &   &    &     &    \\
7                                                       & 7436  & 3471   & 665    & 25   & 13   & 10 & 8 &    &     &    \\
8                                                       & 7401  & 3453    & 708   & 25 & 13   & 10   & 10  & 8  &     &    \\
9                                                       & 7412    & 3459 &  687  & 25    & 13    & 10    & 10  & 8 & 4  &   \\
10                                                       & 7417  & 3450 &  683  & 25    & 13   & 10    & 10  & 8  & 8    & 4 \\* 

\bottomrule
\caption{
\textbf{SPONGE on Menéame Network.} We tested the \textit{SPONGE} algorithm on the network of interactions on Menéame. When we set the number of communities (clusters) to be 1--3, the algorithm over 99\% of all users in one community. When we set the number of clusters to be larger than 3, the algorithm identifies a community with approximately 7000 users and a secondary one containing approximately 4400 users. When the number of clusters increases above 7, the secondary community is further split into two communities of 3400 and 600 users.}

\label{table:sponge-meneame}\\
\end{longtable}

\subsection*{Dense (political) networks}

\subsubsection*{Evaluation on Synthetic Networks}
\begin{figure}[h!]
    \centering
\includegraphics[width=\textwidth]{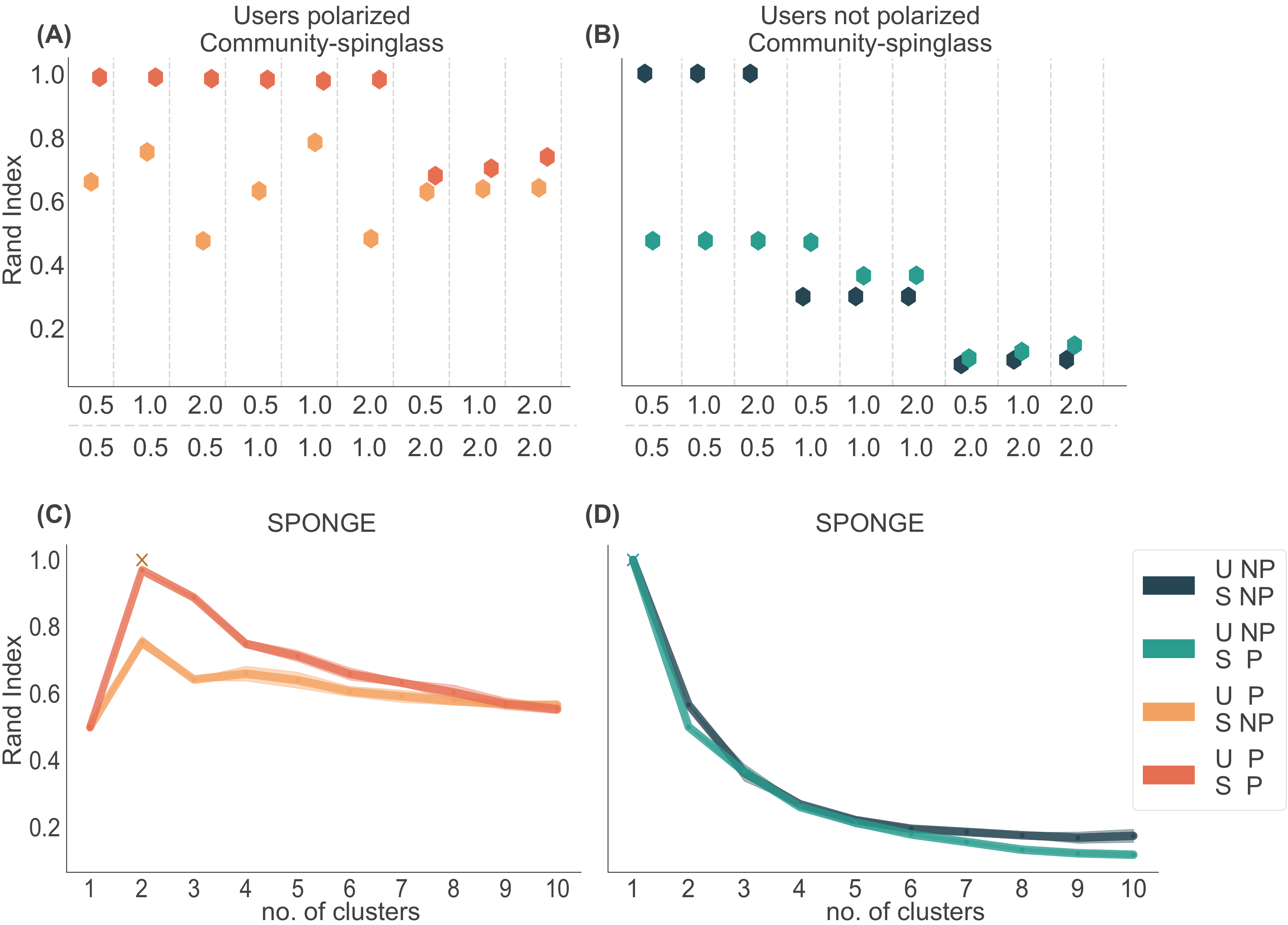}
    \caption{
    \textbf{Community Detection on Dense Synthetic Networks.} 
    We tested the community detection methods on the four synthetic scenarios. Panels (A, C) show the results for the scenarios with two communities of users, while panels (B, D) show the results with one community.  We used the Rand Index to evaluate the performance of the algorithms with different parameter choices.  Higher values of the Rand Index indicate better alignment between expected and empirical communities. Panels (A, B) show the results for \textit{community-spinglass}. We experimented with different combinations of the parameters $\{\gamma^+, \gamma^-\} \in \{0.5, 1, 2\}$. Lower values of $\gamma^+$ indicate less importance given to positive ties in a community, whereas lower values of $\gamma^-$ penalize the presence of negative links within a community. Note that scenarios where users and stories are either polarized or neutral (i.e., \textbf{U NP S NP} and \textbf{U P S P}) are correctly identified for low values of $\gamma^-$. 
    Panels (C, D) show the results for \textit{SPONGE}. We conducted tests for different values of the number of clusters $k$, ranging from $k=1$ (no communities) to $k=10$, and repeated ten times each iteration, due to the stochasticity of the model. Error bars show the standard deviation around the mean. The algorithm correctly identifies the expected communities in scenarios with polarized users (panel C), while it generates spurious communities when users are not polarized (panel D). 
}
    \label{fig:uniform-dense-red}
\end{figure}

Offline voting contexts—such as legislative bodies—are typically dense with few abstentions. They are also characterized by strong partisanship (e.g., legislators voting similarly to others in the same party), which often results in a clear community structure. This structure can simplify the task of community detection. However, we find the opposite to be true. Compared with the sparse case, both \textit{SPONGE} and \textit{community-spinglass} perform worse in all synthetic scenarios (Figure \ref{fig:uniform-dense-red} and Figure \ref{fig:clerk} for the other similarity measures).

Similar to the sparse case, we find that both algorithms perform well when both users and stories are polarized (case \textit{UP SP} in Figure \ref{fig:uniform-dense-red}A,C). Unlike the sparse case, increasing the fixed number of clusters in \textit{SPONGE} decreases the performance (Figure \ref{fig:uniform-dense-red}C). When users are polarized but stories are not polarized, 50--80\% of the user pairs are correctly identified, depending on parameter choice. 

The results of the two algorithms differ when users are expected to be found in a single community (Figure \ref{fig:uniform-dense-red}B,D). \textit{Community-spinglass} algorithm performs well in the case where users and stories are not polarized (for $\gamma^- = 0.5$), and performs poorly when stories are polarized but users are not, irrespectively of parameter choice. \textit{SPONGE} performs extremely poorly for the entire parameter space (except for the trivial case where we force the algorithm to find only one cluster). Increasing the number of clusters results in users being uniformly distributed across all communities. 

\subsubsection*{Evaluation on real networks: US House of Representatives data}
We applied the community detection methods to the US House of Representatives data and considered the political affiliations of the representatives (Democrats, Republicans, and Independents) as the ``true'' communities. 

\begin{figure}[h!]
    \centering
\includegraphics[width=\textwidth]{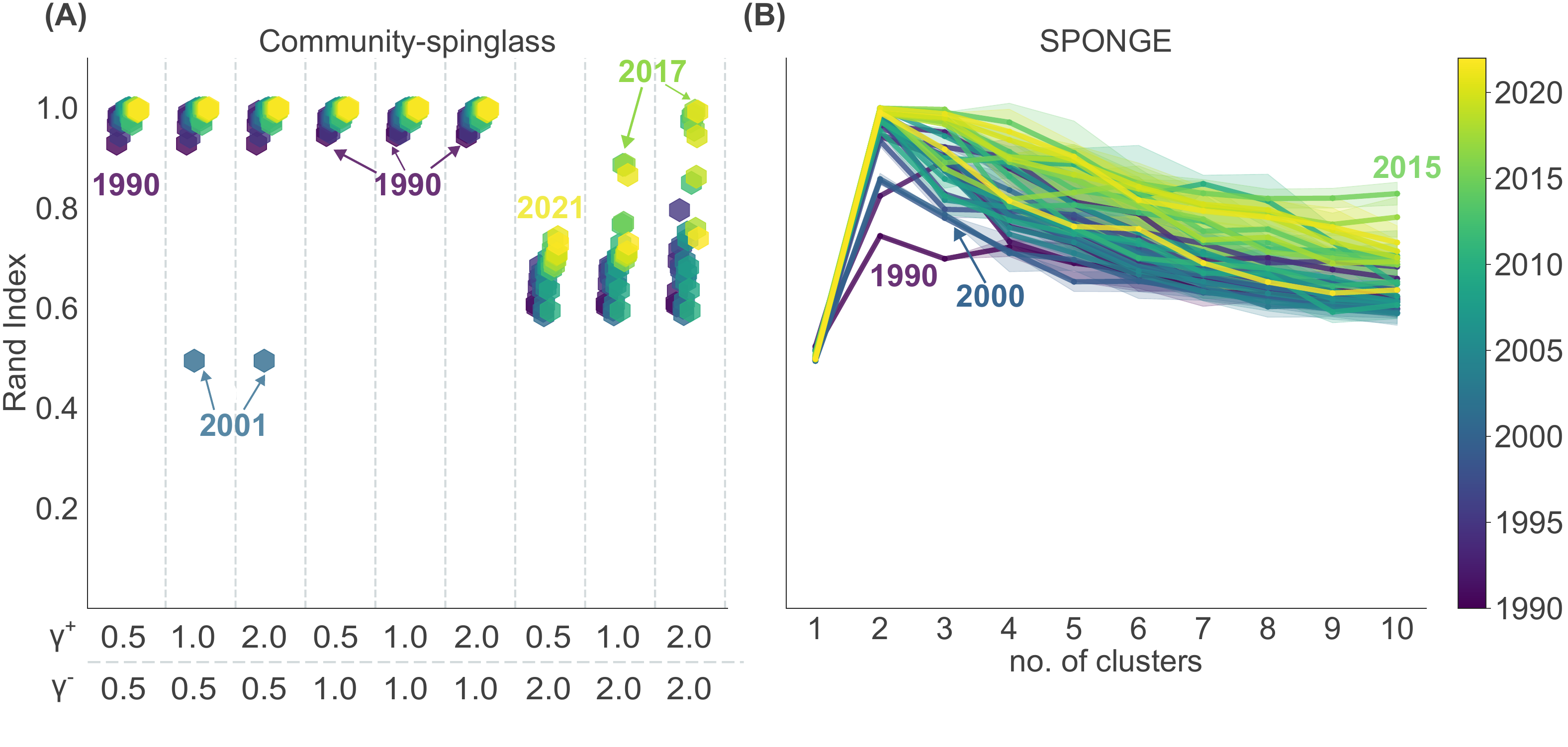}
    \caption{
    \textbf{Community Detection on US House of Representatives Networks.} We tested the community detection methods on 33 co-voting networks resulting from the (dis)agreement of members of the US House of Representatives on several bills, divided per year. We considered the subdivision into political parties (i.e., Democratic, Republican, or Independents) as the ``true'' communities. Panel (A) shows the results for \textit{community-spinglass}.
    We found that parameter choices with $\gamma^- < 2$ consistently capture the subdivision of Representatives into political parties. Panel (B) shows the results for \textit{SPONGE}. Confidence intervals represent the standard deviation range among different runs.
    We conducted tests for different values of the number of clusters $k$, ranging from $k=1$ (no communities) to $k=10$. For $k=1$, we consistently find that the true subdivision is not recovered. However, for subsequent values of $k$, we observe a peak for values $k \in \{2,4\}$ in most of the networks, with a gradual decrease in the method's efficacy as $k$ increases. Note that the Rand Index increased in more recent years.}

    \label{fig:clerk-red}
\end{figure}

We find that \textit{community-spinglass} correctly identifies the political communities by choosing $\gamma^- < 2$ (Figure \ref{fig:clerk-red}A) for all years except 2001. The algorithm is unable to find the political affiliation for 2001 with parameters $\{ \gamma^+, \gamma^- \} = \{ 1, 0.5 \}$ and $\{ \gamma^+, \gamma^- \} = \{ 2, 0.5 \}$. This could be due to the 9/11 attacks. In that year, a large fraction of the legislation\cite{mccartyPolicyEffectsPolitical2007} is attributed to the attacks and received broad support across partisan lines. We find an increasing correlation over time between voting behavior and political affiliation\cite{pooleSpatialModelLegislative1985}--i.e., we are increasingly more able to recover the correct political affiliation from the votes alone. This reflects the increase in partisanship levels through the years observed by Andris et al. \cite{andrisRisePartisanshipSuperCooperators2015} among others.

In the case of \textit{SPONGE} (Figure \ref{fig:clerk-red}B), the political affiliation of the representatives is correctly identified when the number of clusters, $k$, is set to 2--4, with a gradual decrease as $k$ increases. Similar to the \textit{community-spinglass} case, we evidence an increasing ability to recover political affiliations through the years. 

The good recovery of political affiliations for a wide parameter range in both \textit{community-spinglass} and \textit{SPONGE} indicate that both US representatives and US bills are polarized. This is perhaps unsurprising since bills are themselves introduced by US representatives. 

\section*{Conclusion}

In this paper, we systematically assessed the performance of community detection algorithms designed from unipartite signed networks on projected bipartite signed networks, using both synthetic and real-world data. We analyzed sparse networks resembling users' behavior on online social platforms, and dense networks resembling US politicians' votes on bills.

For sparse networks, the algorithms' performance depended on the underlying network structure. Both algorithms were successful at recovering communities of users based on their ideology in the scenarios where two communities of users existed but often identified several communities when only one community of users existed. The results were generally highly dependent on parameter tuning, emphasizing the importance of careful selection in real-world applications.

In the context of dense synthetic networks, both algorithms performed well when both user and story ideologies were polarized, but generated spurious communities in all other scenarios. This was particularly the case for \textit{SPONGE} in the scenarios when only one community of users existed. Both methods well performed on the US House of Representatives data, finding an increasing level of political polarization over time.

In conclusion, our study contributes to the understanding of community detection in bipartite signed networks. The presented benchmark and systematic evaluation provide a foundation for future research in developing robust algorithms for various real-world scenarios. Future studies using community detection in bipartite signed networks can use our framework and open-source code \footnote{\url{https://github.com/elenacandellone/signed-bipartite-nets}} 
to generate synthetic scenarios with realistic voting probabilities. These scenarios will enable researchers to select optimal parameters and evaluate the potential sensitivity of the results to parameter choice. 

Two lessons can be learned from our results. First, the sensitivity of results to parameter choices underscores the importance of robustness tests. Using different parameter choices implies defining communities in different ways. For example, giving more weight to negative ties will create communities without dissent. Giving extra importance to positive ties may help to recover small organized groups. We emphasize the importance of using robustness tests to validate the recovered communities. In particular, external validation (e.g., political affiliation or clustering of user comments in online media) should be used when feasible. 

The second lesson to be learned is the importance of negative ties. For both the dense and sparse network cases, giving extra importance to negative ties in \textit{community-spinglass} resulted in enhanced inference of political affiliations in both sparse and dense networks. This indicates that negative ties are more indicative of communities than positive ties in voting networks. 

The analysis presented has several limitations that open up additional avenues for future research. The first limitation is that we focus on two algorithms for community detection in signed networks frequently used in the literature. Future work could use our benchmark to compare an expanded number of algorithms.

The second limitation is the projection of bipartite networks into unipartite networks, which creates information loss, even though there are only less than 3\% of users with the same number of agreements and disagreements in the real data. At the same time, we test a method that can work directly with the bipartite networks (Stochastic Block Model, see Section \hyperref[subsubsec:sbm]{Weighted SBM with Edge Attributes}, Figure \ref{fig:sbm}), this algorithm is unable to recover the community structure. SBM identifies numerous smaller communities, which, at a higher hierarchical level, merge into a single group (i.e., absence of structure), even in scenarios where two communities are expected. We highlight the need for new methods tailored to bipartite signed networks.

\section*{Author contributions statement}
Conceptualization: All. Data curation: JGB, EC. Formal analysis: EC, SC. Methodology: JGB, EC, EK. Supervision: JGB, EK. Roles/Writing - original draft: EC. Writing - review editing: JGB, EK

\section*{Data and Code Availability}

The data and code associated with this research are publicly available on GitHub. Interested readers can access the repository at the following URL:
\url{https://github.com/elenacandellone/signed-bipartite-nets}.
This repository includes the datasets used in the study, as well as the code implementations of the methods and algorithms discussed in the paper. 


\bibliography{biblio_zotero}
\appendix

\renewcommand\thefigure{A\arabic{figure}}  
\renewcommand\thetable{A\arabic{table}}  

\clearpage
\section*{Appendix}
\subsection*{Ideology difference distribution}\label{appendix:ideology-diff}
Section \hyperref[subsec:synth-nets]{Generating synthetic bipartite signed networks} shows the benchmark we devised to create synthetic networks with simulated voting behavior. To compute voting probabilities, we combined controlled scenarios with real data. In this section, we demonstrate that the distribution of the difference between two random variables is a Gaussian Mixture, under the assumption that the two random variables are also distributed as Gaussian Mixtures.

Take two random variables $X$ and $Y$ distributed as in Equations \ref{eq:ideology_distributions}. Their characteristic functions are given by
\begin{align*}
    \phi_X(t) &= \mathbb{E}\left[ e^{itX}\right] = \int_\mathbb{R} e^{itx} \left[\frac{\pi_x}{\sqrt{2 \pi \sigma_1^2}}\, e^{-\frac{(x-\mu_1)^2}{2\sigma_1^2}} + \frac{\left( 1- \pi_x \right)}{\sqrt{2 \pi \sigma_2^2}}\, e^{-\frac{(x-\mu_2)^2}{2\sigma_2^2}}\right] dx =  \pi_x \cdot e^{-\frac{1}{2}t^2 \sigma_1^2 + it\mu_1} + \left( 1- \pi_x \right) \cdot e^{-\frac{1}{2}t^2 \sigma_2^2 + it\mu_2}, \\
    \phi_Y(t) &= \mathbb{E}\left[ e^{itY}\right] = \int_\mathbb{R} e^{ity} \left[\frac{\pi_y}{\sqrt{2 \pi \sigma_3^2}}\, e^{-\frac{(x-\mu_3)^2}{2\sigma_3^2}} + \frac{\left( 1 - \pi_y \right)}{\sqrt{2 \pi \sigma_4^2}}\, e^{-\frac{(x-\mu_4)^2}{2\sigma_4^2}}\right] dy =  \pi_y \cdot e^{-\frac{1}{2}t^2 \sigma_3^2 + it\mu_3} + \left( 1 - \pi_y \right) \cdot e^{-\frac{1}{2}t^2 \sigma_4^2 + it\mu_4}.
\end{align*}
Then, the random variable $X-Y$ has the following characteristic function 
\begin{align*}
    \phi_{X-Y}(t) &= \mathbb{E}\left[ e^{it(X-Y)}\right] =\mathbb{E}\left[ e^{itX}\right]\mathbb{E}\left[ e^{-itY}\right] =\\
    &=\left[ \pi_x \cdot e^{-\frac{1}{2}t^2 \sigma_1^2 + it\mu_1} + \left( 1- \pi_x \right) \cdot e^{-\frac{1}{2}t^2 \sigma_2^2 + it\mu_2}\right] \cdot 
    \left[ \pi_y \cdot e^{-\frac{1}{2}t^2 \sigma_3^2 - it\mu_3} + \left( 1 - \pi_y \right) \cdot e^{-\frac{1}{2}t^2 \sigma_4^2 - it\mu_4}\right] = \\
    &= \pi_x \cdot \, \pi_y \cdot \left( e^{-\frac{1}{2}t^2 (\sigma_1^2+ \sigma_3^2) + it(\mu_1-\mu_3)}\right) + \pi_x \cdot \, \left( 1 - \pi_y \right) \cdot \left( e^{-\frac{1}{2}t^2 (\sigma_1^2+ \sigma_4^2) + it(\mu_1-\mu_4)}\right) +\\
    &+ \left( 1- \pi_x \right) \cdot \, \pi_y \cdot \left( e^{-\frac{1}{2}t^2 (\sigma_2^2+ \sigma_3^2) + it(\mu_2-\mu_3)}\right) + \left( 1- \pi_x \right) \cdot \, \left( 1 - \pi_y \right) \cdot \left( e^{-\frac{1}{2}t^2 (\sigma_2^2+ \sigma_4^2) + it(\mu_2-\mu_4)}\right),
\end{align*}
that lead to the conclusion that $X-Y$ is distributed as of Equation \ref{eq:ideology-difference}. 

\clearpage
\subsection*{Degree-corrected synthetic networks}\label{appendix:deg-corr}
To generate more realistic synthetic bipartite signed networks, we can extend the procedure explained in Subsection \hyperref[subsec:synth-nets]{Generating synthetic bipartite signed networks}. Instead of using fixed voting probabilities for all users, we can employ a voting probability distribution. This implies that we can replace step 3 in the previous mechanism with a range of probabilities, as follows:
\begin{enumerate}
    \item Extract from data the positive and negative voting distributions;
    \item Assign randomly to each user a positive and negative ``degree'', keeping the same coupling as in the data;
    \item Repeat the voting procedure as before, but when the user reaches their assigned ``degree'', stop assigning votes.
\end{enumerate}
With this modification, we aim to preserve the skewness of the voting distribution. In other words, we sample a realistic voting behavior from the data. It is particularly meaningful in the case of sparse networks, where a few users vote for almost every story, while most users will barely vote a few times.

\begin{figure}[h!]
    \centering
\includegraphics[width=\textwidth]{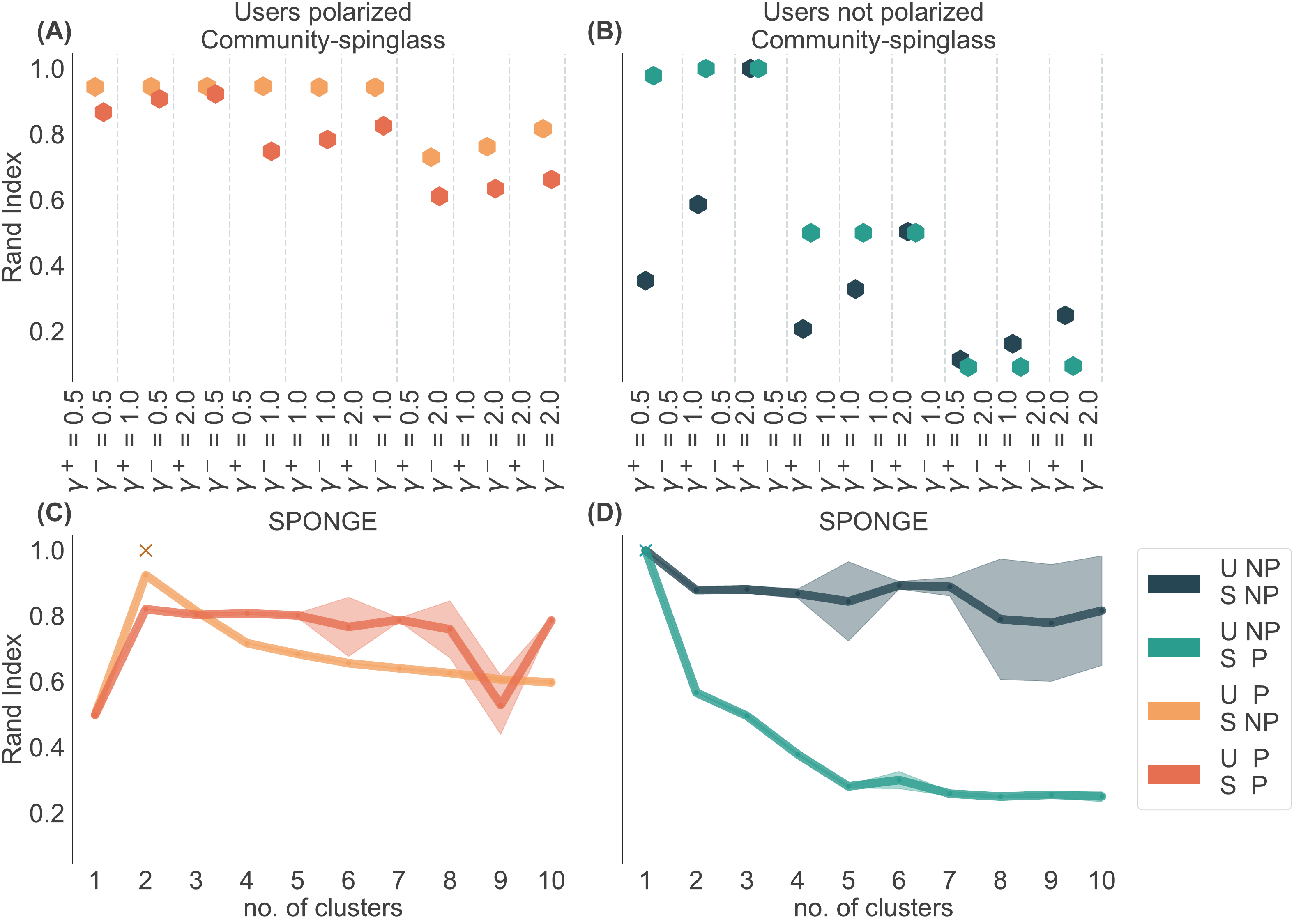}
    \caption{
    \textbf{Community Detection on Menéame Synthetic Networks, degree-corrected version.} We tested the community detection methods on the four synthetic scenarios, with voting probabilities sampled from sparse data's voting distributions. Panels (A)-(B) show the results for \textit{community-spinglass}. Lower values of $\gamma^+$ indicate less importance given to positive ties in a community, whereas lower values of $\gamma^-$ penalize the presence of negative links within a community.
     Note that $\gamma^- = 0.5$ is generally the best parameter choice, as it finds the expected communities for the synthetic scenarios, except for the case where users and stories aren't polarized. Panels (C)-(D) show the results for \textit{SPONGE}. We observed that the algorithm correctly identifies the expected communities in scenarios with polarized users (panel C), while it generates spurious communities in cases where stories introduce latent ideologies (\textbf{U NP S P} case).
   }
    \label{fig:deg-corr}
\end{figure}

Figure \ref{fig:deg-corr} displays the results for sparse networks. Similar to the case of uniform voting probabilities, \textit{community-spinglass} better captures the expected communities for lower values of $\gamma^-,$ except for the scenario where both users and stories aren't polarized (\textbf{U NP S NP}). One possible explanation lies in the increased complexity of the model, as users exhibit similar voting behaviors compared to real data, making them less identifiable within a neutral, unique group. However, \textit{SPONGE} outperforms all scenarios, identifying approximately $80\%$ of the user pairs in the correct clusters, except for the scenario \textbf{U NP S P}, where increasing the number of clusters leads to a decrease in accuracy. We observe that, when introducing additional complexity to the model, community detection methods perform similarly or even worse than in the uniform voting case.

\clearpage
\subsection*{Clustering Evaluation Metrics}\label{appendix:eval-metrics}
Section \hyperref[subsec:eval-measures]{Evaluating Community Similarity} explained the metric used to evaluate the similarity between the expected communities, both in real data with ground truth and in synthetic networks, to the empirical ones found by the community detection methods. However, the Rand Index is not the only metric available for clustering evaluation. It is possible to correct the RI for random assignments, and it is called the \textit{Adjusted Rand Index} \cite{hubertComparingPartitions1985a}, defined as:

\begin{equation}
    \text{ARI} = \frac{\text{RI} - E[\text{RI}]}{\max(\text{RI}) - E[\text{RI}]}.
\end{equation}

Another metric implemented in the \textit{scikit-learn} package is based on the information-theoretic concept of Mutual Information (MI) \cite{vinhInformationTheoreticMeasures2009}. The MI is defined as

\begin{equation}
    \text{MI}(\mathbf{\sigma_T}, \mathbf{\sigma_E}) = \sum_{i=1}^{|\mathbf{\sigma_T}|} \sum_{j=1}^{|\mathbf{\sigma_E}|} \frac{|\sigma_T^i \cap \sigma_E^j|}{N}\log\left(\frac{N|\sigma_T^i \cap \sigma_E^j|}{|\sigma_T^i||\sigma_E^j|}\right),
\end{equation}

and can be normalized, obtaining the so-called Normalized Mutual Information,

\begin{equation}
    \text{NMI}(\mathbf{\sigma_T}, \mathbf{\sigma_E}) = \frac{\text{MI}(\mathbf{\sigma_T}, \mathbf{\sigma_E})}{\text{mean}(H(\mathbf{\sigma_T}), H(\mathbf{\sigma_E}))},
\end{equation}

where $\mathbf{\sigma_T}$ and $\mathbf{\sigma_E}$ are respectively the true and empirical configurations. Another possible clustering evaluation measure is called the v-score \cite{rosenbergVMeasureConditionalEntropyBased2007}. It is defined as the harmonic mean of two functions, the homogeneity $h$ and the completeness $c$, defined as

\begin{equation*}
\begin{split}
  h &= 1 - \frac{H(C|K)}{H(C)}\\
  c &= 1 - \frac{H(K|C)}{H(K)}.
\end{split}
\end{equation*}

\begin{figure}[h!]
    \centering
\includegraphics[width=\textwidth]{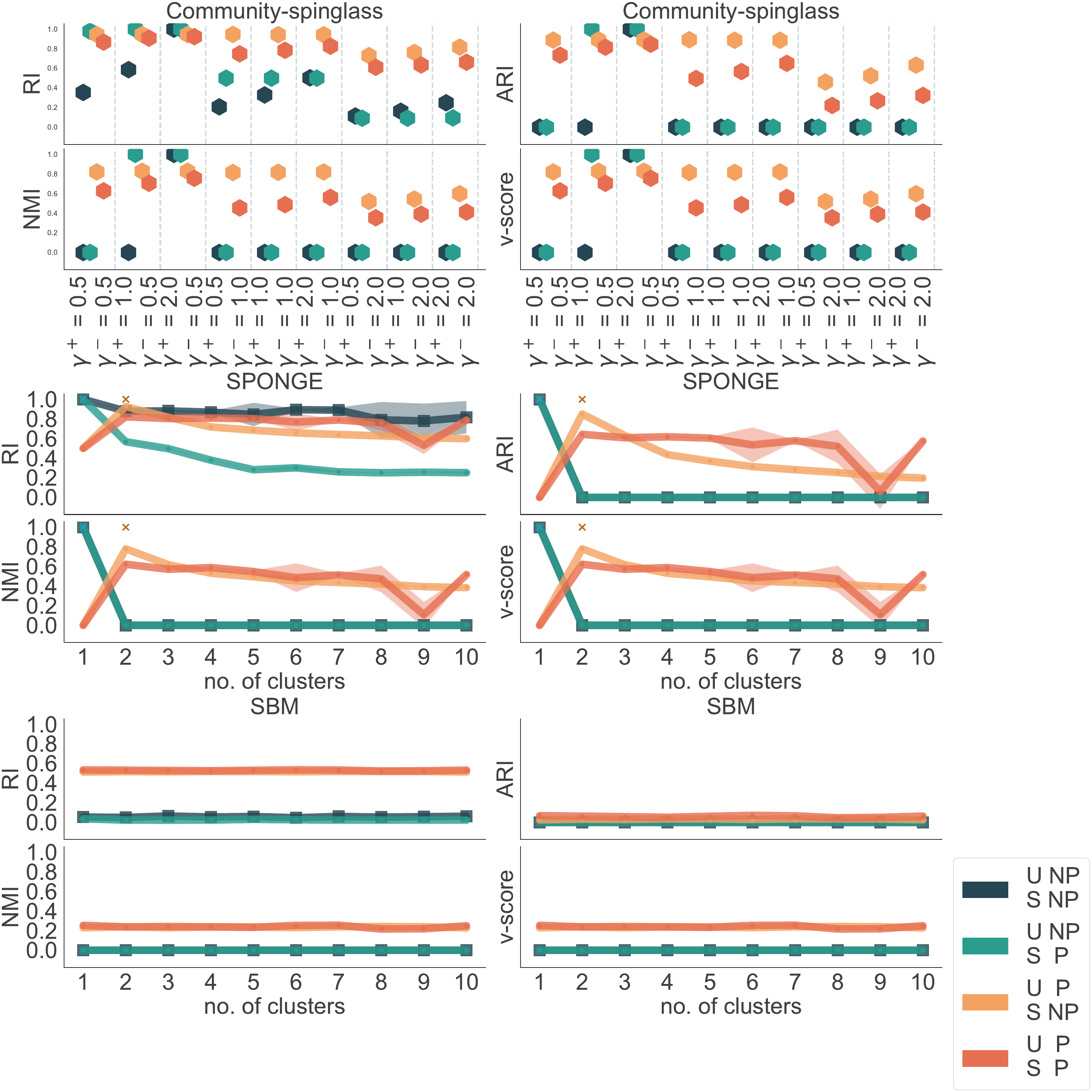}
    \caption{
    \textbf{Community Detection on Menéame Synthetic Networks, comparing several evaluation metrics.} We conducted tests on four synthetic scenarios to evaluate community detection methods. We used four evaluation metrics: Rand Index (RI) in panels (A)-(E), Adjusted Rand Index (ARI) in panels (B)-(F), Normalized Mutual Information (NMI) in panels (C)-(G), and v-score in panels (D)-(H). We compared the results of \textit{community-spinglass} and \textit{SPONGE}. We found that the last three metrics showed similar behavior in failing to identify cases where one community was expected.
}
    \label{fig:uniform}
\end{figure}

\begin{figure}[h!]
    \centering
\includegraphics[width=\textwidth]{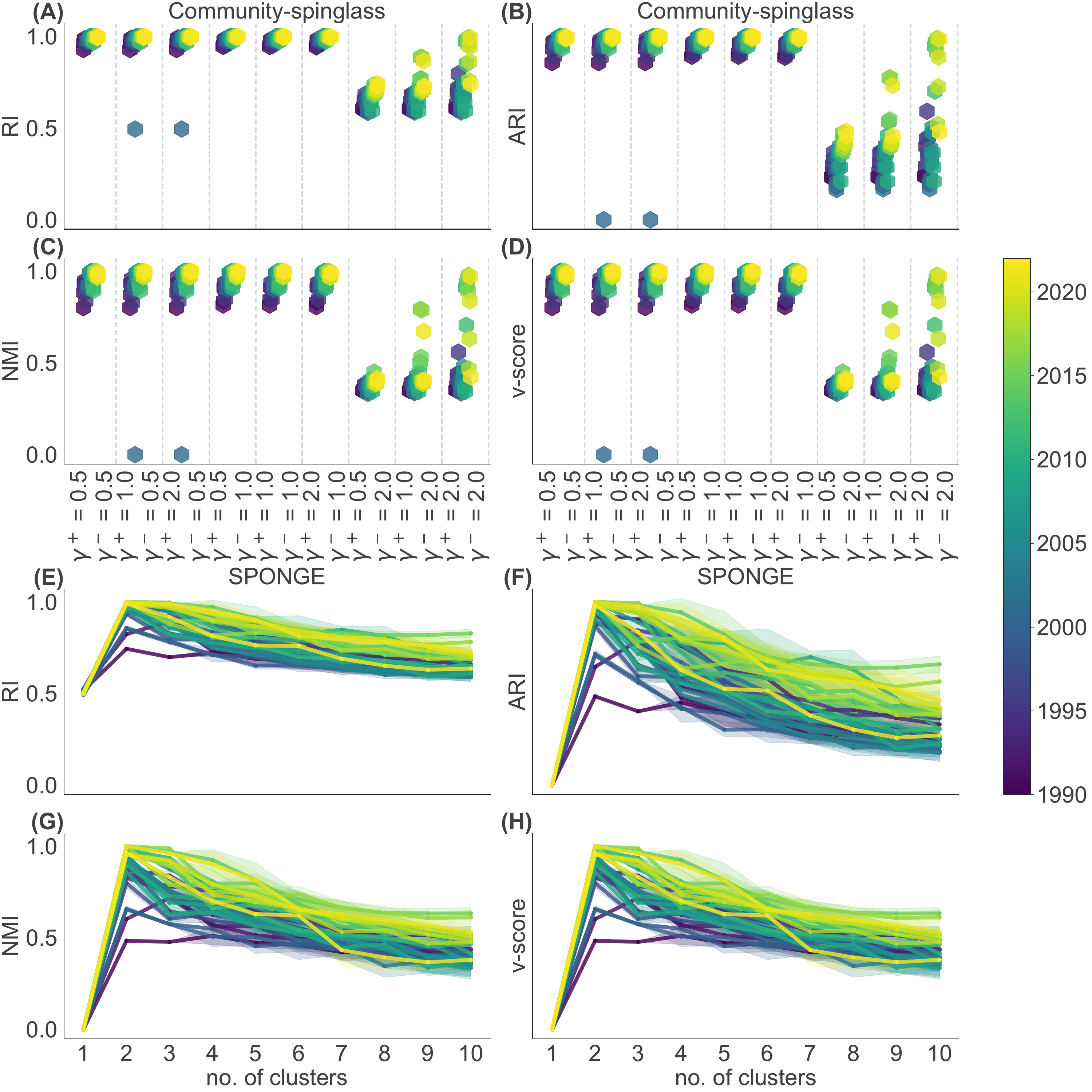}    
\caption{
    \textbf{Community Detection on US House of Representatives Networks, various evaluation metrics.} We tested the community detection methods on the US House of Representatives data. We compare four evaluation metrics: Rand Index (RI) in panels (A)-(E), Adjusted Rand Index (ARI) in panels (B)-(F), Normalized Mutual Information (NMI) in panels (C)-(G), and v-score in panels (D)-(H). We show the results for \textit{community-spinglass} and \textit{SPONGE}. As we identified the ``true'' subdivision as a tripartition in political affiliations, we found a coherent pattern among metrics. }

    \label{fig:clerk}
\end{figure}

Figure \ref{fig:uniform} complements the results presented in Figure \ref{fig:uniform-red} by including the other evaluation metrics explained in this section. We show that metrics such as the ARI, NMI, and v-score, while generally preferable due to their adjustments for randomness, may fail to evaluate cases where we expect to find a single community. We find a coherent pattern in the case of the US House of Representatives Networks (Figure \ref{fig:clerk}), where we chose as expected communities the political affiliations (i.e., Democrats, Republicans, and Independent), therefore the four metrics highlight the same behavior as the Rand Index.

\clearpage
\subsection*{Descriptive statistics of co-voting networks of US House of Representatives}\label{appendix:descr-stat}
\begin{longtable}[c]{@{}ccclccc@{}}
\toprule
year & number of nodes & number of edges & | & year & number of nodes & number of edges \\* \midrule
\endhead
\bottomrule
\endfoot
\endlastfoot
1990 & 435 & 94169 &  & 2007 & 444 & 98202 \\
1991 & 441 & 96716 &  & 2008 & 445 & 97982 \\
1992 & 435 & 94232 &  & 2009 & 447 & 99444 \\
1993 & 443 & 97315 &  & 2010 & 447 & 99543 \\
1994 & 442 & 97290 &  & 2011 & 440 & 96530 \\
1995 & 443 & 97787 &  & 2012 & 439 & 95854 \\
1996 & 439 & 95844 &  & 2013 & 443 & 97541 \\
1997 & 439 & 95334 &  & 2014 & 438 & 95527 \\
1998 & 439 & 95485 &  & 2015 & 445 & 98239 \\
1999 & 439 & 95833 &  & 2016 & 438 & 95437 \\
2000 & 436 & 94671 &  & 2017 & 443 & 97331 \\
2001 & 444 & 98122 &  & 2018 & 442 & 96777 \\
2002 & 437 & 94981 &  & 2019 & 449 & 100026 \\
2003 & 436 & 94603 &  & 2020 & 443 & 97293 \\
2004 & 438 & 94977 &  & 2021 & 441 & 96943 \\
2005 & 438 & 95464 &  & 2022 & 449 & 99752 \\
2006 & 437 & 95035 &  & \multicolumn{1}{l}{} & \multicolumn{1}{l}{} & \multicolumn{1}{l}{} \\* \bottomrule
\caption{\textbf{Snapshot co-voting networks of US House of Representatives.} We collected data from 1990 to 2022 of the House of Representatives votes. We then subdivided the dataset per year and we generated the bipartite network of Representatives and Bills into (dis)agreement unipartite networks where the nodes are the Representatives and the edges are the co-votings between two Representatives. The weight of those edges is given by Equation \ref{eq:weight-projection}.}
\label{tab:clerk-descr-stats}\\
\end{longtable}

\clearpage
\subsection*{Weighted SBM with Edge Attributes}\label{subsubsec:sbm}

The Stochastic Block Model (SBM) \cite{hollandStochasticBlockmodelsFirst1983} is a generative process for creating networks that exhibit a partition into blocks. This model is highly flexible, allowing not only the generation of networks with a predefined block structure but also the inference of communities in existing networks \cite{peixotoBayesianStochasticBlockmodeling2019b}.
Given our context of weighted signed networks, we implemented the weighted, degree-corrected version of the SBM \cite{peixotoNonparametricWeightedStochastic2018}, incorporating the sign as an additional edge attribute. We utilized the SBM version in the Python package \textit{graph-tool} \cite{peixotoGraphtoolPythonLibrary2017}. As suggested in the package documentation, we sampled the logarithm of the absolute value of edge weights from a normal distribution and the edge sign, rescaled to values between 0 and 1 (i.e., 0 corresponds to the negative sign, 1 to the positive sign), from a Bernoulli distribution. In contrast with previous methods, frustration minimization is not the explicit mechanism to generate communities. Therefore, it would also be possible to find communities with maximal frustration, with only negative edges inside a community and positive between communities.

\begin{figure}[h!]
    \centering
\includegraphics[width=\textwidth]{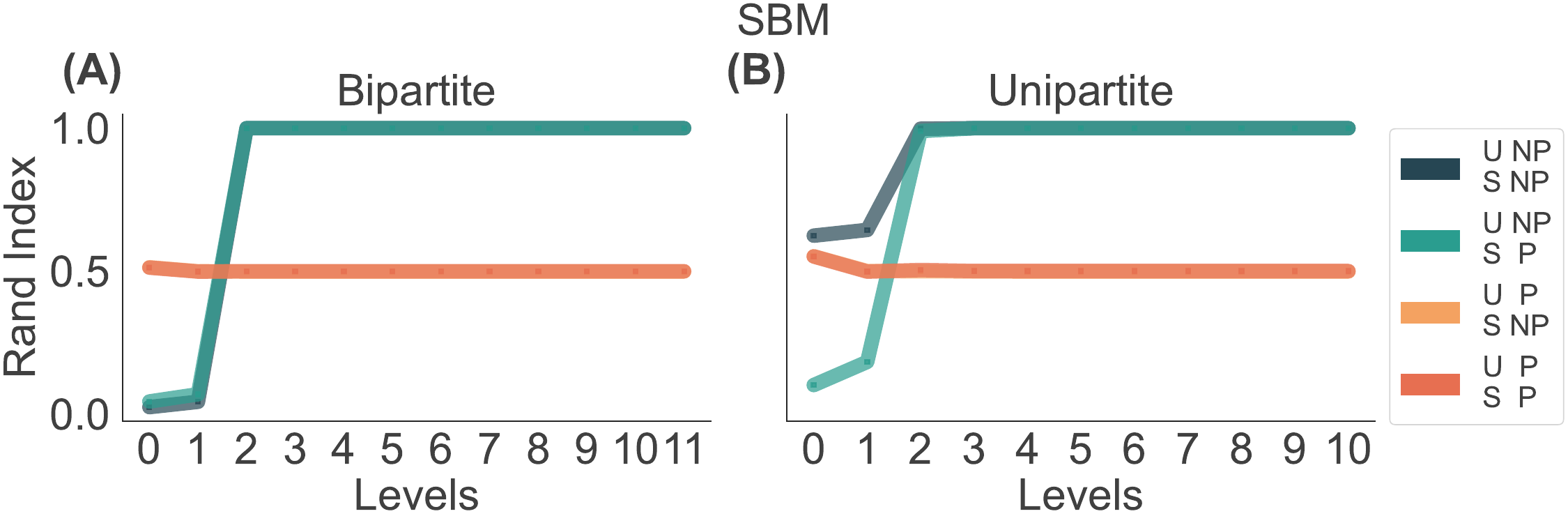}
    \caption{
    \textbf{Stochastic Block Model on Sparse Synthetic Networks.} We applied the hierarchical SBM to four sparse synthetic scenarios. Panel (A) displays the results on the bipartite network, i.e., the user-story network without projecting to a user-user network. An additional node attribute, the set membership, is given as a parameter of the algorithm. Panel (B) illustrates the results of the bipartite projection. We observe that in both cases, after two hierarchical levels, all four scenarios collapse into a unique community. This results in the maximum Rand Index (RI) for the \textbf{U NP} cases and approximately $50\%$ for the other two cases. However, examining the lower hierarchical levels reveals that the SBM identifies a high number of communities for all scenarios, in contrast to the other algorithms analyzed in this paper.
}
    \label{fig:sbm}
\end{figure}

Figure \ref{fig:sbm} depicts the results obtained by applying SBM to both the bipartite network, without the projection (with the option to add set membership as a node attribute), and the unipartite, projected networks. In both cases, we employed a hierarchical SBM.
However, in both cases, the algorithm does not correctly identify the expected communities. Nonetheless, this is not necessarily a limitation, as the algorithm is not constrained to a specific optimization problem like previous algorithms. Instead, at the lower hierarchical levels it detects numerous smaller groups that may represent different optimal scenarios compared to the expected ones.

\clearpage
\subsection*{Comparing algorithms on real data}\label{appendix:comparison-real-data-communities}
\begin{figure}[h!]
    \centering
    \includegraphics[width=\textwidth]{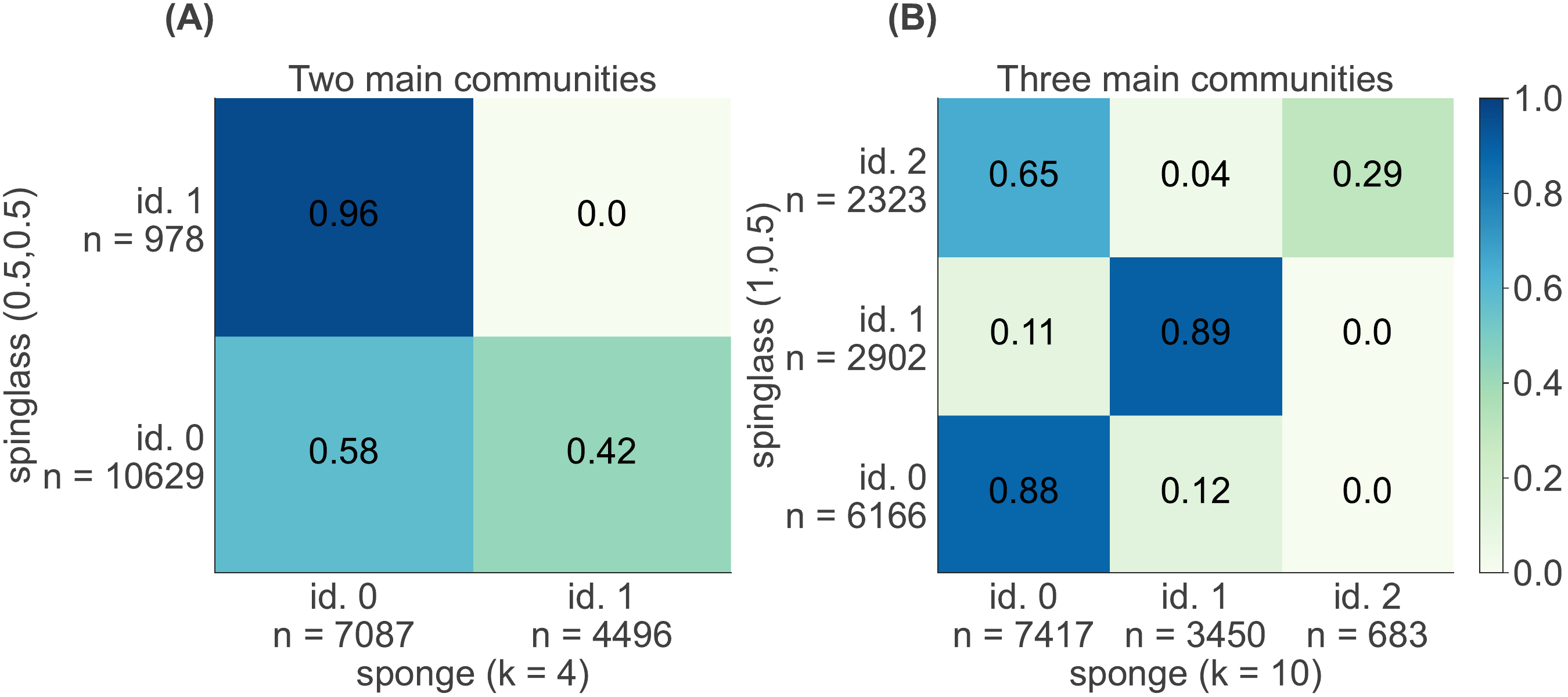}
    \caption{\textbf{Comparing methods results on Menéame data.} We found that when using specific parameter choices, both community detection methods recover either two or three main communities. Panel (A) displays the results for parameter choices that yield two larger communities for both algorithms. It is normalized by row and smaller communities are not shown. We observed that 60\% of community 0 in \textit{community-spinglass} is shared with community 0 in \textit{SPONGE}, with the remaining 40\% found in community 1 of SPONGE. Panel (B) shows the results for parameter configurations presenting three main communities. Communities 0 and 1 from \textit{community-spinglass} can be identified respectively as communities 0 and 1 from \textit{SPONGE} with an accuracy of around 88-89\%.}
    \label{fig:methods-comparison-meneame-real}
\end{figure}

Results subsection \hyperref[subsubsec:meneame-data-results]{Menéame data} shows that, depending on the choice of the parameters, both methods recover either two or three main communities in the Menéame Network. To further investigate the accuracy of these results, we examine whether the primary communities identified by each algorithm were the same or not. Figure \ref{fig:methods-comparison-meneame-real} displays two distinct parameter choices that generate either two or three main communities. We observe that in the former case, communities 0 and 1 contain 60\% and 40\% of the largest community respectively, as detected by \textit{SPONGE}. In the latter case, both \textit{SPONGE} and \textit{community-spinglass} accurately detected communities 0 and 1 with an accuracy of around 88-89\%. These results show that although the methods follow the same principle of frustration minimization, they produce varying results and do not consistently generate stable communities for all parameter choices.

\end{document}